\documentclass[twocolumn,pra,superscriptaddress]{revtex4-2}
\usepackage{graphicx}
\usepackage{amsmath}
\usepackage{amssymb}
\usepackage{xcolor}
\usepackage{subfigure}
\usepackage{bm}
\usepackage{dsfont}
\usepackage{hyperref}
\usepackage{todonotes}
\usepackage{physics}

\begin{document}
	
	\title{Geometric decomposition of geodesics and null phase curves using Majorana star representation}
	\author{Vikash Mittal}
	\email{vikashmittal.iiser@gmail.com}
	\affiliation{Department of Physical Sciences, Indian Institute of Science Education \& Research (IISER) Mohali, Sector 81 SAS Nagar, Manauli PO 140306, Punjab, India}
	
	\author{Akhilesh K.~S.}
	\email{akhileshks@iisermohali.ac.in}
	\affiliation{Department of Physical Sciences, Indian Institute of Science Education \& Research (IISER) Mohali, Sector 81 SAS Nagar, Manauli PO 140306, Punjab, India}
	
	\author{Sandeep K.~Goyal}
	\email{skgoyal@iisermohali.ac.in}
	\affiliation{Department of Physical Sciences, Indian Institute of Science Education \& Research (IISER) Mohali, Sector 81 SAS Nagar, Manauli PO 140306, Punjab, India}
	
	\begin{abstract}
		Geodesics are the shortest curves between any two points on a given surface. Geodesics in the state space of quantum systems play an important role in the theory of geometric phases, as these are also the curves along which the acquired geometric phase is zero. Null-phase curves (NPCs) are the generalization of the geodesics, which are defined as the curves along which the acquired geometric phase is zero even though they need not be the shortest curves between two points. Here we present a geometric decomposition of geodesics and NPCs in higher-dimensional state space, which allows understanding of the intrinsic symmetries of these curves. We use Majorana star representation to decompose a geodesic in the $n$-dimensional Hilbert space to $n-1$ curves on the Bloch sphere and show that all the $n-1$ curves are circular segments with specific properties that are determined by the inner product of the end states connected by the given geodesic. We also propose a method to construct infinitely many NPCs between any two arbitrary states for $(n>2)$-dimensional Hilbert space using our geometric decomposition.

	\end{abstract}
	
	\maketitle
	
	\section{Introduction}
	A geodesic is the shortest path between two points on a given surface. In the state space of a quantum system, the geodesics are also the curves along which the acquired geometric phase is zero~\cite{Samuel1988,Mukunda1993}. Hence, they play a crucial role in the theory of geometric phases. Further, they are also used in designing optimal quantum circuits, which turned out to be equivalent to finding the shortest path between two points in a certain curved geometry~\cite{Nielsen2006}. Geodesics can be generalized to a larger class of curves, known as null-phase curves (NPCs)~\cite{Rabei1999}. An NPC is defined as a curve between two pure quantum states on the state space along which the acquired geometric phase is zero~\cite{Arvind2003,Chaturvedi2013}. Unlike geodesics, the NPCs need not be the shortest path between the two states. The role of the geometric phase in characterizing topological phases of matter~\cite{Zak1989}, in precision measurements~\cite{Martinez2011,HuYu2012,Arya2022}, and in robust quantum information processing~\cite{Jones2000,Vedral2003} highlights the importance of understanding NPCs and geodesics. In this paper, we present a geometric decomposition of the geodesics and NPCs for the higher-dimensional quantum systems to a set of curves on the Bloch sphere. This decomposition reveals the hidden symmetries of higher-dimensional geodesics and NPCs and may facilitate a deeper understanding of the state space structure for such systems.

	To find the geometric decomposition of the geodesics and NPCs, we use the Majorana star (MS) representation, which enables the representation of a state of an $n$-level quantum system by a symmetric combination of  $(n-1)$ states of a two-level systems~\cite{Majorana1932, BlochRabi1945}. This representation has found applications in quantum information~\cite{Solano2009,Bastin2010,Dogra2018}, quantum entanglement~\cite{Messina2010, Makela2010, Markham2011, Devi2012, Fu2016, Aulbach2011}, geometric phases~\cite{Hannay1998, Fu2014, Fu2016, Akhilesh2020}, and topological phases of matter~\cite{Chen2015, Gong2020}. Recently, the bulk topology and the bulk-boundary correspondence have been studied in the non-Hermitian tight-binding model using the MS representation~\cite{Zhao2021}.
	
	Since in the MS representation, a state $\ket{\Psi}$ of an $n$-level quantum system can be mapped to a symmetric state of $n-1$ number of two-level systems, it can be represented by a set of $n-1$ points on the Bloch sphere. Hence, a curve on the state space of an $n$-level system can be mapped to $n-1$ curves on the Bloch sphere. The $n-1$ points corresponding to the state $\ket{\Psi}$ are often called MSs, and the collection of the points is referred to as a constellation.
	
	In this paper, we decompose the geodesics and NPCs of higher-dimensional quantum systems into curves on the Bloch sphere using MS representation. The key findings of this paper are the following: (i) geodesics of the $n$-level quantum system decompose to $n-1$ circular segments on the Bloch sphere when the end states are represented by $(n-1)$-fold degenerate MSs. When $n$ is odd, the $n-1$ curves occur in pairs that are reflection of each other about the great circle on the Bloch sphere connecting the end states. For even $n$, one curve is along the great circle connecting the end states, and the remaining $n-2$ curves occur in pairs reflective about the same great circle. (ii) For odd $n$, a class of NPCs can be constructed using $(n-1)/2$ pairs of curves which are reflective about a great circle, whereas, for even $n$,  $(n-2)/2$ pairs of curves are reflective, and the remaining curve can be chosen along the great circle connecting the two states. 
	
	Our treatment provides a deeper understanding and inherent symmetries of geodesics in higher-dimensional state space. For example, geodesics in three-dimensional state space, where the end states are chosen such that each one of them is represented by degenerate MSs, decompose into two curves. We found that these two curves together form a unique circle on the Bloch sphere, where the end states are the diagonally opposite points on the circle. The radius of the circle depends solely on the inner product between the end states. Therefore, we can generate a geodesic between two states in three-dimensional state space by constructing a circle on the Bloch sphere between the corresponding end points. Since any pair of three-dimensional states can be mapped to states represented by degenerate MSs using a unitary transformation, we can construct a geodesic between any arbitrary states. 
	
	Using our geometric decomposition,  we construct a prominent class of NPCs for $(n>2)$-dimensional state space. These NPCs can be constructed by choosing curves in pairs such that the curves within a pair are reflections of each other. If the total number of curves is odd, then one curve can be chosen along the great circle connecting the end states on the Bloch sphere. Since there exists an infinite number of such pairs between any two end states, we can construct infinitely many NPCs. A special subset of these NPCs is where the curves are reflections of themselves, i.e., all the curves are along the great circle connecting the end points. This subset can be of experimental importance while designing quantum circuits.

	This paper is organized as follows: In Sec.~\ref{sec:Background}, we discuss the relevant background required for our main results. Here, we discuss  MS representation, geodesics, and the NPCs. The geometric decomposition of geodesics is given in Sec.~\ref{Sec:Geodesic}, and of NPCs is given in Sec.~\ref{Sec:NPC}. We conclude in Sec~\ref{sec:Conclusion}.
	
	\section{Background}\label{sec:Background}
	In this section, we present the relevant background necessary for our main results. Here, we introduce the MS representation for $n$-dimensional quantum systems. We discuss the  connection between Bargmann invariant (BI) and geometric phases and end this section by providing the definitions of geodesics and NPCs.

	\subsection{MS representation} \label{subsec:MSR}
	Symmetric subspace of $n-1$ two-level quantum systems is $n$ dimensional which is isomorphic to an $n$-level quantum system. Hence, the state of an $n$-level system can be geometrically represented as a collection of $n-1$ points on the Bloch sphere, which is known as MS representation~\cite{Majorana1932,BlochRabi1945}. In this section, we briefly outline the details of MS representation.
	
	Consider a general $n$-level state $\ket{\Psi}$ written as
	\begin{equation} \label{eq:spin-j state}
		\ket{\Psi}=\sum_{r=0}^{n-1}c_{r}\ket{r},
	\end{equation}
	where $c_{r}$ are the expansion coefficients such that $\sum_r|c_r|^2= 1$, and $\{\ket{r}\}$ is the computational basis. The same state $\ket{\Psi}$ can also be written as a symmetric superposition of $n-1$ two-level systems as
	\begin{equation} 
		\ket*{\tilde{\Psi}} = \mathcal{N} \sum_{P}\big[ \ket{\psi_1} \otimes \ket{\psi_2}\otimes \dots \otimes \ket{\psi_{n-1}} \big]. \label{Eq:Sym}
	\end{equation}
	Here, $\sum_{P}$ corresponds to the sum over all $(n-1)!$ permutations of the qubits, and $\mathcal{N}$ is the normalization factor. From here onward, we denote the state in the MS representation with a tilde sign on the state, i.e., a state $\ket{\Psi}$ in the normal representation will read $\ket*{\tilde\Psi}$ in MS representation.

	The state $\ket{\psi_k} = \alpha_k \ket{0} +\beta_k\ket{1}$ represents a state of a two-level system. To arrive at the MS representation, $\ket{\psi_k}$ is expressed in dual-rail representation~\cite{Milburn2007}, i.e., $\ket{\psi_k} \equiv (\alpha_ka_1^\dagger + \beta_ka_2^\dagger)\ket{0,0}$, where $a^{\dagger}_1$, $a^{\dagger}_2$ are the bosonic creation operators for two independent modes, and $\ket{0,0}$ is the two-mode vacuum state. The symmetrized state of $n-1$ two-level systems in this representation can simply be written as $\prod_{k=1}^{n-1}(\alpha_k a^{\dagger}_1+\beta_ka^{\dagger}_2)\ket{0,0}$ due to the  indistinguishable nature of $n-1$  bosons. Hence,  
	\begin{align} \label{eq:dual-ray-qubit}
		\ket*{\tilde{\Psi}} &\equiv \prod_{k=1}^{n-1}(\alpha_k a^{\dagger}_1+\beta_ka^{\dagger}_2)\ket{0,0},\\
		&\equiv\sum_{r=0}^{n-1}c_r\dfrac{(a_1^\dagger)^{n-1-r}(a^{\dagger}_2)^{r}}{\sqrt{r! (n-1-r)!}}\ket{0,0}.
	\end{align}
	Comparing Eqs.~\eqref{eq:spin-j state} and~\eqref{eq:dual-ray-qubit}, we get
	\begin{equation}
		\ket{r} =\dfrac{(a^{\dagger}_1)^{n-1-r}(a^{\dagger}_2)^{r}}{\sqrt{r! (n-1-r)!}}\ket{0,0},
	\end{equation}
	and coefficients $c_r$ are functions of $\alpha_k$ and $\beta_k$. Now the task is to evaluate the $\alpha_k$ and $\beta_k$ from given  $c_r$. This can be achieved by constructing a polynomial of the form:
	\begin{equation}
		\prod_{k=1}^{n-1}(\alpha_k x - \beta_k) \equiv \sum_{r=0}^{n-1} f_{r} x^{n-1-r} = 0 \label{eq:Majorana-polynomial}
	\end{equation}
	where
	\begin{equation}\label{eq:Majorana-coefficients}
		f_{r} = (-1)^{r} \dfrac{c_{r}}{\sqrt{r! (n-1-r)!}}.
	\end{equation}
	Hence, solving the polynomial in Eq.~\eqref{eq:Majorana-polynomial} yields $\alpha_k$ and $\beta_k$.
	
	\subsection{BI and geometric phase }           
	In this subsection, we define the BI and its relation with the geometric phase. Given three non-orthogonal states $\{\ket{\Psi_1}, \ket{\Psi_2}, \ket{\Psi_3}\}$ from the Hilbert space $\mathcal{H}$, i.e., $\ip{\Psi_i}{\Psi_j}\ne 0; \forall ~i\ne j$, the BI of third order is defined as
	\begin{equation} \label{third-order-BI}
		\Delta_3(\Psi_1,\Psi_2,\Psi_3)= \ip*{\Psi_1}{\Psi_2}\ip*{\Psi_2}{\Psi_3}\ip*{\Psi_3}{\Psi_1}.
	\end{equation}     
	The BI is invariant under unitary transformation $\ket{\Psi_i} \to U\ket{\Psi_i}$ and it plays a crucial role in the theory of geometric phase. Consider a closed curve $\mathcal{C}$ constructed by connecting the three nonorthogonal states $\{\ket{\Psi_i}\}$ by geodesics, then the geometric phase $\Phi_{\text{g}}$ associated with this closed curve is  given by~\cite{Mukunda1993}
	\begin{equation}
		\label{BI-geom}
		\Phi_{\text{g}}[\mathcal{C}]=-\arg\Delta_3(\Psi_1,\Psi_2,\Psi_3).
	\end{equation}
	It is straightforward to generalize Eq.~\eqref{third-order-BI} to define $n$th order BI as
	\begin{align} \label{eq:nth-order-BI}
		\Delta_n(\Psi_1,\dots,\Psi_n) = \ip*{\Psi_1}{\Psi_2}\ip*{\Psi_2}{\Psi_3} \dots\ip*{\Psi_n}{\Psi_1}.
	\end{align}
	Further, any higher-order BIs can be reduced to third-order BIs~\cite{Mukunda1993}, and correspondingly, the geometric phase for a closed curve constructed by connecting $n$ states via geodesics can be expressed as the sum of geometric phases for $n-2$ third-order BIs.
	
	\subsection{Geodesic curves} 
	A geodesic is the path of shortest distance between two points on a surface. In this subsection, we introduce a differential equation for the geodesic in the state space of a quantum system. 
	To define geodesic curves, we need a continuously parameterized smooth curve $\mathcal{C}$ in the Hilbert space  $\mathcal{H}$ given by
	\begin{equation}
		\label{g24}
		\mathcal{C}=\{\ket{\Psi(s)}\in \mathcal{H}\;|\;s_1\le s\le s_2\},
	\end{equation}
	where $s$ is the real parameter varied over $[s_1,s_2]$. 
	The quantity called length associated with $\mathcal{C}$ is defined as~\cite{Mukunda1993}
	\begin{equation} \label{length}
		\mathcal{L}  = \int_{s_1}^{s_2}ds \ip*{u_{\perp}(s)}^{1/2},
	\end{equation}
	where
	\begin{equation*}
		\ket{u_{\perp}(s)} = \ket{u(s)}-\ip*{\Psi(s)}{u(s)}\ket{\Psi(s)},
	\end{equation*}
	and $\ket{u(s)}$ is the tangent to $\ket{\Psi(s)}$ which reads
	\begin{equation*}
		\ket{u(s)} = \frac{d}{ds} \ket{\Psi(s)} \equiv \ket*{\dot{\Psi}(s)}.
	\end{equation*}
	By requiring $\delta{\cal{L}}=0$, we obtain a differential equation obeyed by $\mathcal{C}$ to be a geodesic:
	\begin{equation} \label{eq:diffgeodesic}
		\left[\frac{d}{ds}-\ip*{\Psi(s)}{u(s)}\right]\frac{\ket{u_{\perp}(s)}}{\norm{u_{\perp}(s)}}=f(s)\ket{\Psi(s)},
	\end{equation} 
	where $f(s)$ is a real function of $s$ which is yet to be determined.
	
	The geodesics are invariant under the $U(1)$ transformation of the form $\ket{\Psi(s)} \to e^{i\alpha(s)} \ket{\Psi(s)}$. By exploiting this freedom and the freedom of reparameterization, we can make $\ip{\Psi(s)}{u(s)} = 0$ and $ \norm{u} = $ constant, respectively, which yields~\cite{Mukunda1993,Arvind2003}
	\begin{equation}
		\dfrac{d^2}{ds^2} \ket{\Psi(s)} = f(s)\ket{\Psi(s)}.
	\end{equation}
	Further, the above equation can be shown equivalent to a differential equation of the form~\cite{Mukunda1993}:
	\begin{align} \label{reduced-geodesic-eq}
		\dfrac{d^2}{ds^2} \ket{\Psi(s)} = -\ip*{\dot{\Psi}(s)}{\dot{\Psi}(s)}\ket{\Psi(s)}.
	\end{align}
	A formal solution of Eq.~\eqref{reduced-geodesic-eq}, for the given two end states $\ket{\Psi(s_1)}$ and $\ket{\Psi(s_2)}$ with $\ip*{\Psi(s_1)}{\Psi(s_2)} \equiv \xi$ real and positive, is given by
	\begin{equation}
		\label{geodesic-curve}
		\ket{\Psi(s) } = \cos(s)\ket{\Psi(s_1)}+\frac{\ket{\Psi(s_2)}-\xi\ket{\Psi(s_1)}}{(1-\xi^2)^{1/2}}\sin(s).
	\end{equation}
	Hence, using Eq.~\eqref{geodesic-curve}, we can generate the geodesic between any two points in the state space.
	
	\subsection{NPCs} 
	NPCs are the curves between two points on the quantum state space along which the acquired geometric phase is zero~\cite{Rabei1999,Arvind2003,Chaturvedi2013}. Mathematically, these curves can be defined as follows: Consider a differentiable curve $\{\ket{\Psi(s)}\}$ for the real parameter $s \in (s_1,s_2)$ such that $\ip{\Psi(s)}{\Psi(s')}\ne 0 $ for all $s,s'\in(s_1,s_2)$. 
	The curve $\{\ket{\Psi(s)}\}$ is an  NPC if, for any three points  on the  curve, the BI is real and positive, i.e.,
	\begin{equation} \label{null-phase-def-1}
		\Delta_3 \left[\Psi(s), \Psi(s'), \Psi(s'')\right] > 0 , \; s,s',s''\in [s_1,s_2].
	\end{equation}
	From the above definition, it is clear that, if the curve $\{\ket{\Psi(s)}\}$ is an NPC, then $e^{i \beta} \{\ket{\Psi(s)}\}$ will also be an NPC. Exploiting this condition, we can always choose a curve in the $\mathcal{H}$ such that
	\begin{equation}
		\label{NPC in unit sphere}
		\ip{\Psi(s)}{\Psi(s')} > 0,
	\end{equation}
	for any $s,s'\in(s_1,s_2)$~\cite{Arvind2003}. Hence, there exist infinitely many NPCs between any two points in the state space.

	\section{Bloch sphere decomposition of geodesics} \label{Sec:Geodesic}
	The geodesic between any two states of a two-level quantum system is the segment of the great circle connecting these states on the Bloch sphere. This is the consequence of the spherical geometry of the state space of a two-level system. However, the geodesics in three- or higher-dimensional state spaces are notoriously difficult to understand, even though the expression to calculate these geodesics is given in Eq.~\eqref{geodesic-curve}. 
	In this section, we present the Bloch sphere decomposition of higher-dimensional geodesics using MS representation. This Bloch sphere decomposition reveals intrinsic symmetries of geodesics which may help to understand the geometric structure of higher-dimensional  state space. We start with geodesics in three-dimensional state space and extend these results to higher dimensions.
	
	\subsection{Geodesics in three-dimensional state space} \label{sec:geodesics3}
	Consider two states $\{\ket{\Psi_1}, \ket{\Psi_2} \}$ in the three-dimensional state space. For simplicity, we choose states  of the following form:
	\begin{equation} \label{eq:endstates}
		\ket{\Psi_1} = \begin{pmatrix}
			1\\
			0\\
			0
		\end{pmatrix},\qquad \ket{\Psi_2} = \begin{pmatrix}
			\alpha^2 \\
			\sqrt{2} \alpha \beta \\
			\beta^2
		\end{pmatrix},
	\end{equation}
	such that, each one of them, individually, is represented by degenerate MSs. Here, $\alpha$ and $\beta$ are real, with $ \ip*{\Psi_1}{\Psi_2} = \alpha^2 \equiv \cos\theta$, $ 0 \le \theta < \pi/2$, and $\alpha^2 + \beta^2 = 1$.
	
	In the MS representation, the states considered in Eq.~\eqref{eq:endstates} take the form:
	\begin{eqnarray}
		\ket*{\tilde{\Psi}_1} &=& \ket{0} \otimes \ket{0},\nonumber\\
		\ket*{\tilde{\Psi}_2} &=& \ket{\phi} \otimes \ket{\phi}, \label{Eq:End-states}
	\end{eqnarray} where $ \ket{0} = \begin{pmatrix} 1 & 0\end{pmatrix}^T$ and $ \ket{\phi} = \begin{pmatrix} \alpha & \beta \end{pmatrix}^T$. 
	
	From Eq.\eqref{geodesic-curve}, we can find a geodesic $ \{\ket{\Psi(s)} \;|\; 0 \le s \le \theta \} $ connecting $\ket{\Psi_1}$ and $\ket{\Psi_2}$  which reads
	\begin{align} \label{eq:geodesic}
		\ket{\Psi'(s)} &= \begin{pmatrix}
			\cos (s) \\
			a \sin (s) \\
			b \sin (s)
		\end{pmatrix},
	\end{align}
	which further can be written in MS representation as 
	\begin{align}
		\ket*{\tilde\Psi(s)}\equiv\dfrac{1}{\mathcal{N}(s)}\left[\ket{\psi_+(s)}\ket{\psi_-(s)}+\ket{\psi_-(s)}\ket{\psi_+(s)}\right].
	\end{align}
	Here, $a=\sqrt{2} \alpha \beta/\sin \theta $, $b= \beta^2/\sin \theta$, and $\mathcal{N}(s)$ is the normalization constant. The states $\ket{\psi_\pm}$ in the MS representation of $\ket{\Psi(s)}$ are given by 
	\begin{equation} 
		\label{Majorana-decomposition}
		\ket{\psi_{\pm}(s)} = \dfrac{1}{\sqrt{1 + \abs*{x_{\pm}(s)}^2}} \begin{pmatrix}
			1 \\
			x_{\pm}(s)
		\end{pmatrix},
	\end{equation}
	where  
	\begin{equation}
		\label{solutions}
		x_{\pm}(s) = \dfrac{a\sin(s)\pm i \sqrt{b\sin(2s)-a^2\sin^2(s)}}{\sqrt{2}\cos(s)}.
	\end{equation}
	Here, $x_{\pm}(s)$ are the solutions of the Majorana polynomial in Eq.~\eqref{eq:Majorana-polynomial} corresponding to the $\ket{\Psi(s)}$ in  Eq.~\eqref{eq:geodesic}.
	Thus, the geodesic curve $\{\ket{\Psi(s)}\}$ decomposes into two curves $ \{\ket{\psi_+(s)} \;|\; 0 \le s \le \theta \} $ and $ \{\ket{\psi_-(s)} \;|\; 0 \le s \le \theta \} $ belonging to  two-dimensional space. Note that the curves $ \{\ket{\psi_{\pm}(s)} \;|\; 0 \le s \le \theta \} $ themselves do not satisfy the differential equation for a geodesic in Eq.~\eqref{eq:diffgeodesic}. Therefore, they are not geodesics in two-dimensional space.
	
	The Bloch vectors corresponding to  $ \{\ket{\psi_{\pm}(s)\}}  $  are denoted as $\{\vb{n}_{\pm}(s)\}$ and obtained by
	\begin{equation}
		\label{components on the Bloch sphere}
		\vb{n}_{\pm}(s) = \ev{\boldsymbol{\sigma}}{\psi_{\pm}(s)},
	\end{equation}
	where $\boldsymbol{\sigma} = \{\sigma_x, \sigma_y, \sigma_z\}$ is the vector of Pauli matrices~\cite{Nielsen2010}. 
	The components of the curves $\{\vb{n}_{\pm}(s)\}$ along $x,y,z$ read
	\begin{eqnarray}
		\label{reflection-bloch-sphere}
		\left(\vb{n}_{+}\right)_x&=&\left(\vb{n}_{-}\right)_x=\dfrac{\sqrt{2}a\sin(s)}{\cos(s)+b\sin(s)}, \nonumber\\
		\left(\vb{n}_{+}\right)_y&=&-\left(\vb{n}_{-}\right)_y=\dfrac{\sqrt{b\sin(2s)-a^2\sin^2(s)}}{\cos(s)+b\sin(s)}, \nonumber\\
		\left(\vb{n}_{+}\right)_z&=&\left(\vb{n}_{-}\right)_z=\dfrac{\cos(s)-b\sin(s)}{\cos(s)+b\sin(s)}.
	\end{eqnarray} 
	\begin{figure}
		\subfigure[]{
			\includegraphics[width=3.5cm]{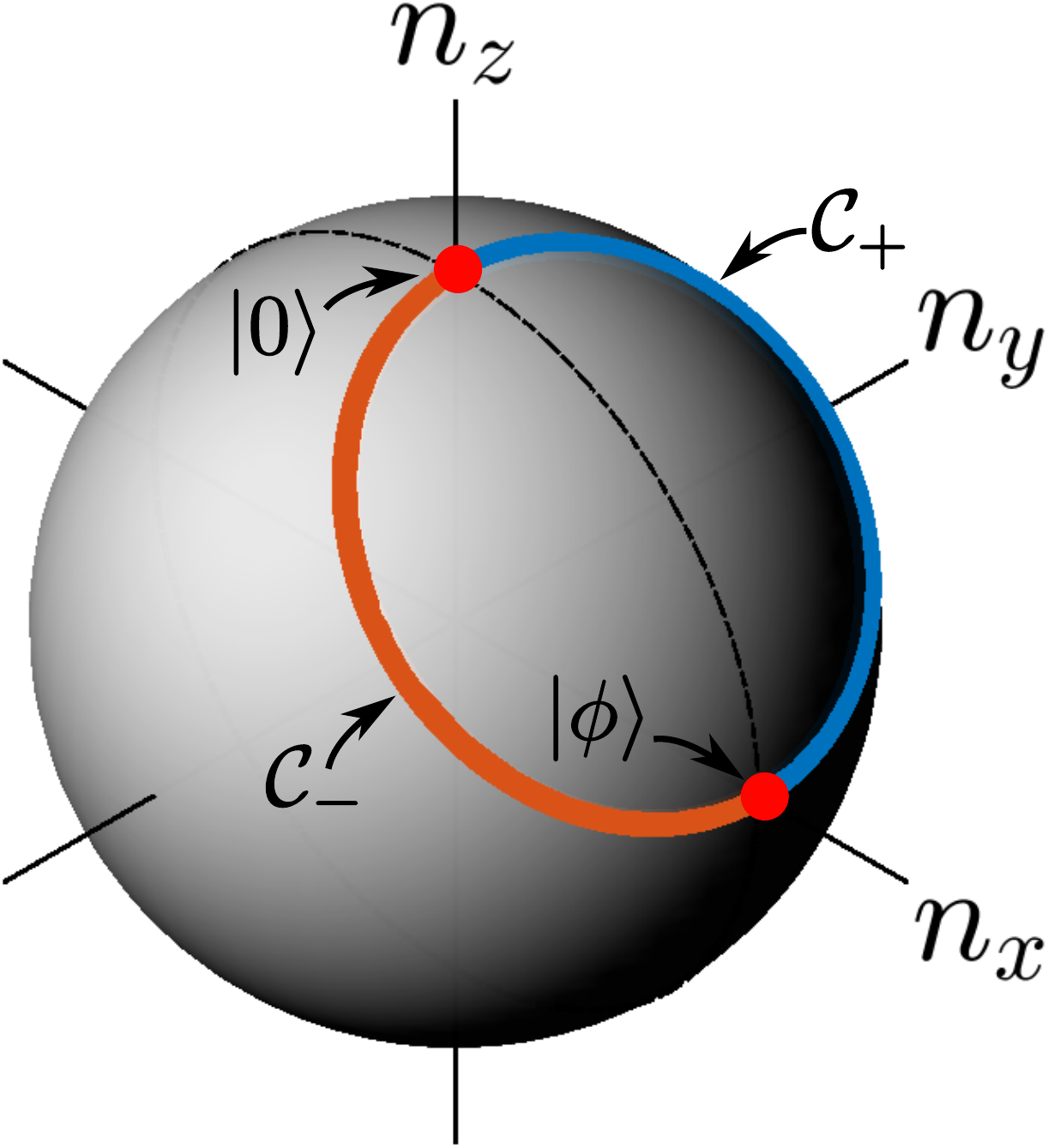}
			\label{fig:geod1}}
		\subfigure[]{
			\includegraphics[width=3.5cm]{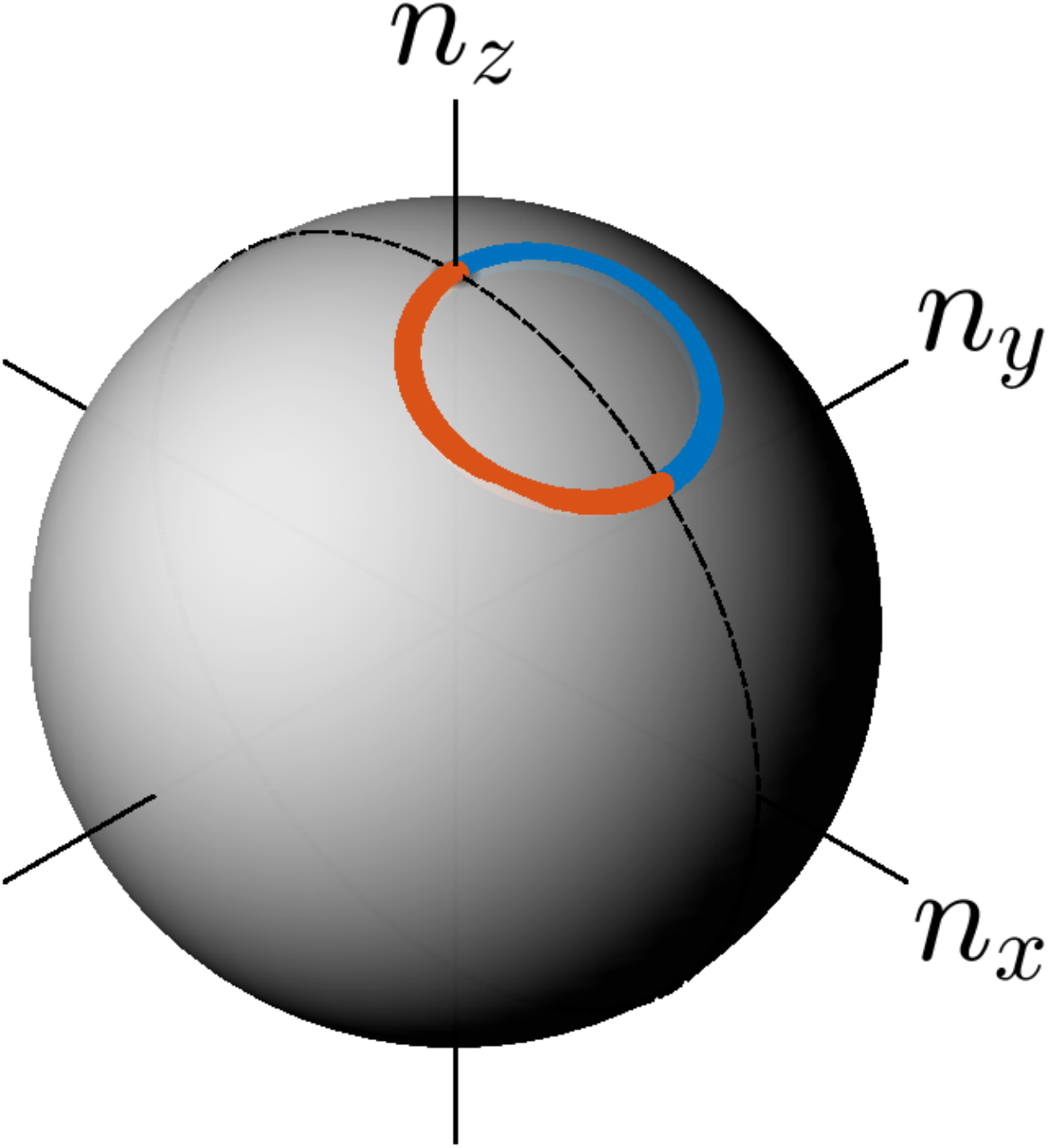}
			\label{fig:geod2}}
		\caption{(Color online) Here we plot the geometric decomposition of a geodesic between two states given in Eq.~\eqref{Eq:End-states} where we have chosen \subref{fig:geod1} $\theta = \pi/3$, \subref{fig:geod2} $\theta = \pi/5$. The blue and orange curves correspond to $\{\vb{n}_+\}$ and $\{\vb{n}_-\}$, respectively, given in Eq. \eqref{reflection-bloch-sphere}. }
		\label{fig:geodesic}
	\end{figure}
	Since  the end states $\ket{0}$ and $\ket{\phi}$ in Eq.~\eqref{Eq:End-states} lie on the $xz$ plane on the Bloch sphere and the solutions $\{x_\pm(s)\}$ in Eq.~\eqref{solutions} form a complex conjugate pair, the components of the curves $\{\vb{n}_{\pm}(s)\}$ differ only along the $ y $ axis by a negative sign. Therefore, the two curves are reflective about the $xz$ plane. We will call these curves \emph{dual} of each other.
	
	On a careful observation,  we can see that the components of the pair of dual curves along the $x,y,z$ axes satisfy an equation of the form
	\begin{equation}
		\label{eq-of-circle}
		\left[(\vb{n}_{\pm})_x-\alpha\beta\right]^2+\left(\vb{n}_{\pm}\right)_y^2+\left[(\vb{n}_{\pm})_z-\alpha^2\right]^2=\beta^2,
	\end{equation}
	which is an equation of a circle with center at $(\alpha\beta,0,\alpha^2)$, the midpoint of the line joining the end states on the  Bloch sphere. Hence, the two curves are segments of a circle and share the same center and the same radius $\beta$. Therefore, the geodesic curve connecting two states in the three-dimensional state space can be identified by a complete circle of radius $\beta =\sqrt{1-\ip{\Psi_1}{\Psi_2}}$ on the Bloch sphere constructed by joining two semicircular arcs $\{\vb{n}_\pm(s)\}$.
	The  pair of dual curves  $\mathcal{C}_{\pm} \equiv \{ \vb{n}_{\pm}(s)\}$ traced by the two states in Eq.~\eqref{Majorana-decomposition} are shown in Fig. \ref{fig:geodesic} for different values of $\theta$.

	So far, we have considered the geodesics between the end states which are represented by degenerate MSs given in Eq.~\eqref{eq:endstates}. Interestingly, this formalism can be applied to the geodesics between two  arbitrary  three-dimensional states.
	We note that any two states $\{\ket{\Psi_1},\ket{\Psi_2}\}$  in three-dimensional Hilbert space can be transformed to degenerate MSs  by a common unitary transformation $U$ (see Appendix~\ref{AppendixA}). Therefore, we can study the structure of the geodesics between these states by first mapping them to degenerate MSs and constructing the geodesic using the semicircular curves $\mathcal{C}_\pm$. Applying $U^\dagger$ on these curves will result the in actual curves corresponding to the geodesic between $\{\ket{\Psi_1},\ket{\Psi_2}\}$.
	
	To summarize  the results obtained in this section: (i) A geodesic connecting the  three-dimensional states $\ket{\Psi_1}$ and $\ket{\Psi_2}$, which are represented by degenerate MSs on the Bloch sphere,  decomposes into two unique curves on the Bloch sphere.
	(ii) These two curves are reflective about the great circle connecting the two end points on the Bloch sphere  and constitute a circle of radius  $\sqrt{1 -\ip*{\Psi_1}{\Psi_2}}$.
	(iii) One can obtain the geodesic connecting any two arbitrary states in three-dimensional space by first converting the two end states to degenerate MSs states by using a unitary transformation and then constructing the unique circle between the end states on the Bloch sphere.

	\subsection{Geodesics in higher-dimensional state space}
	We extend our analysis to study the structure of geodesics in higher-dimensional state space. Let us start by considering nonorthogonal end states $ \{\ket{\Psi_1}, \ket{\Psi_2}\} $ in an $n$-dimensional state space which map to $(n-1)$-fold degenerate MSs, individually, on the Bloch sphere. In the MS representation, the end states can be written as
	\begin{align}\label{Eq:end-state}
		\ket*{\tilde{\Psi}_1} &= \ket{0}_0 \otimes \ket{0}_1 \otimes \dots \otimes \ket{0}_{n-2}, \nonumber \\
		\ket*{\tilde{\Psi}_2} &= \ket{\phi}_0 \otimes \ket{\phi}_1\otimes \dots \otimes \ket{\phi}_{n-2},
	\end{align}
	where $ \ket{0} = \begin{pmatrix} 1 & 0\end{pmatrix}^T$ and $ \ket{\phi} = \begin{pmatrix} \alpha & \beta \end{pmatrix}^T$. Here, $\alpha$ and $\beta$ are real, with $ \ip*{\Psi_1}{\Psi_2} = \alpha^{n-1} \equiv \cos\theta$, $ 0 \le \theta < \pi/2$, and $\alpha^2 + \beta^2 = 1$.
	From Eq.~\eqref{geodesic-curve}, the geodesic curve  $\{\ket{\Psi(s)};0\le s\le \theta\}$ connecting the end states in the MS representation turns out to be
	\begin{align} \label{eq:PsiTilde}
		\ket*{\tilde\Psi(s)} &= \ket{0}^{\otimes n - 1} + A^{(n-1)}(s) \ket{\phi}^{\otimes n-1}\\
		&\equiv \mathcal{N} \sum_{P} [\ket{\chi_1(s)} \otimes \ket{\chi_2(s)} \otimes \dots \otimes \ket{\chi_{n-1}(s)}],
	\end{align}
	where $\sum_{P}$ corresponds to the sum over all symmetric permutations of the $n-1$ states $ \ket{\chi_k(s)} $  of two-level systems.  Here, $\mathcal{N}$ is the normalization constant, and 
	\begin{align}
		A^{(n-1)}(s) = \dfrac{\sin s}{\cos s (1 - \alpha^{n-1})^{1/2} - \alpha^{n-1} \sin s}.
	\end{align}
	Since the end states considered in Eq.~\eqref{Eq:end-state} are real, the Majorana polynomial [Eq.~\eqref{eq:Majorana-polynomial}] is also real. Therefore,  the roots occur as complex conjugate pairs along with a real root depending on the dimension of the state space. By solving the Majorana polynomial in Eq.~\eqref{eq:Majorana-polynomial}, one can find the curves $\{\ket{\chi_k(s)}\}$, $i = 1, \dots , n-1$ as
	\begin{align}\label{eq:ansatz}
		\ket{\chi_k(s)} = \ket{0} + \Delta\, \omega_k A(s)\ket{\phi}.
	\end{align}
	Here, $\omega_k$ are the $(n-1)$th roots of unity given by $\omega_k = \exp\left(2 \pi i k/(n-1)\right)$ with $k = 0,1,\dots, n-2$, and $\Delta = \left(\prod_{k=0}^{n-2} \omega_k\right)^{1/(n-1)}$. This shows that a geodesic curve in an $n$-dimensional state space decomposes to $n-1$ curves on the Bloch sphere.

	The Bloch vector corresponding to the state $\{\ket{\chi_k(s)}\}$ can be written as 
	$\{\vb{n}_{k}(s) = \ev{\boldsymbol{\sigma}}{\chi_{k}(s)}\}$.  Next, we show that the curve  $\{\ket{\chi_k(s)}\}$ traced by the states $\ket{\chi_k(s)}$ for all the values of $k$ constitute circular segments on the Bloch sphere.
	
	Consider three Bloch vectors $ \vb{p_1},~ \vb{p_2},~ \vb{p_3}$  corresponding to three states on the curve $\{\ket{\chi_k(s)}\}$. The unit vector $\vb{m}$ orthogonal to the plane  containing these three Bloch vectors  can be written as 
	\begin{equation} \label{normal-vector}
		\vb{m}=\dfrac{(\vb{p}_2-\vb{p}_1) \times (\vb{p}_3-\vb{p}_1)}{\norm{(\vb{p}_2-\vb{p}_1)\times (\vb{p}_3-\vb{p}_1)}}.
	\end{equation}
	The intersection of the plane containing the three vectors $ \vb{p_1},~ \vb{p_2},~ \vb{p_3}$ and the Bloch sphere constitute a circle. The three Bloch vectors $\vb{p}_1,~\vb{p}_2,~\vb{p}_3$ also lie on the circle whereas the unit vector $\vb{m}$ passes through the center of this circle. Therefore, the projection of any of these three Bloch vectors will be the same on the vector $\vb{m}$.
	
\begin{figure}
		\subfigure[]{
			\includegraphics[width=3.5cm]{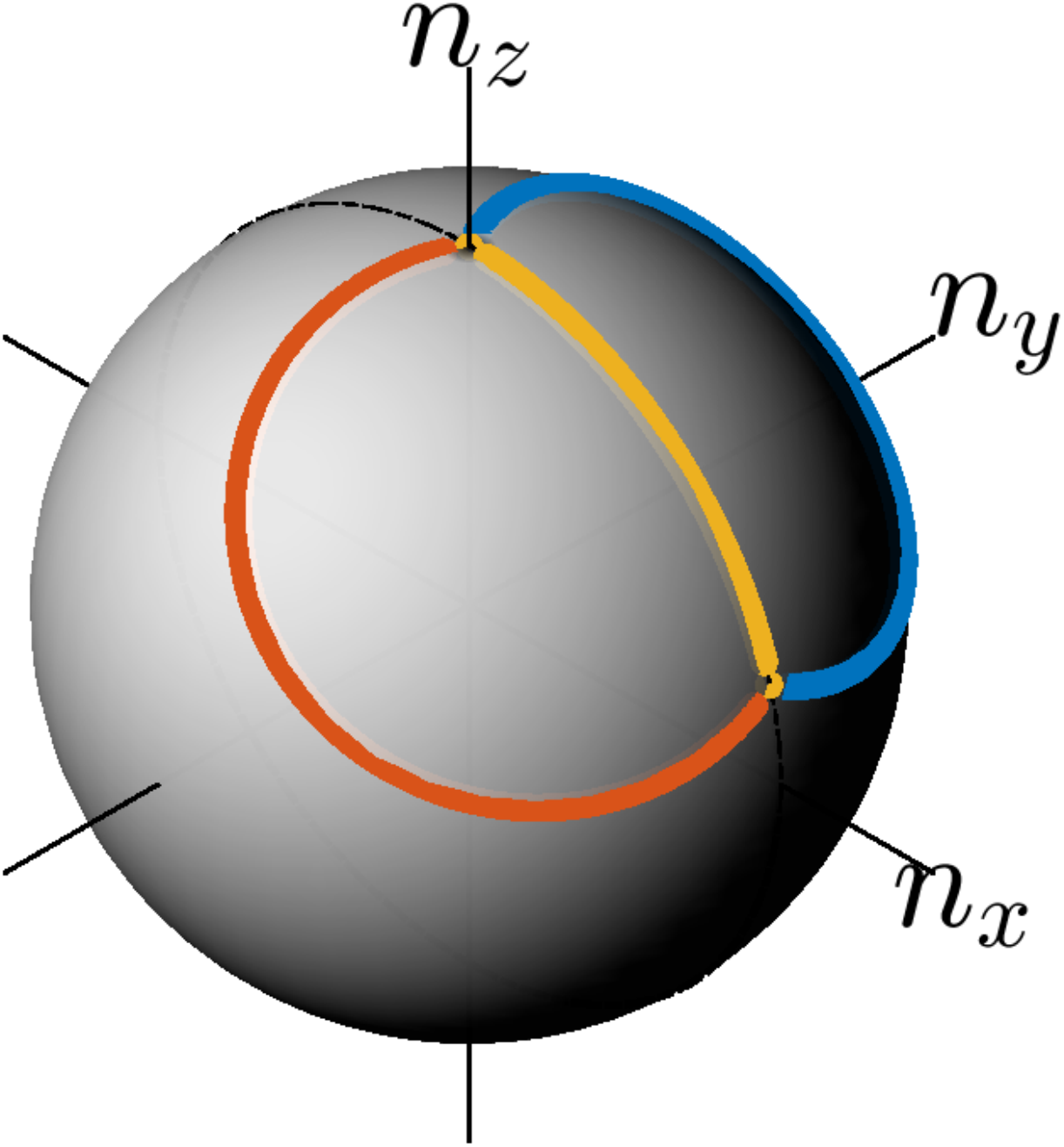}
			\label{fig:geo4D}}
		\subfigure[]{
			\includegraphics[width=3.5cm]{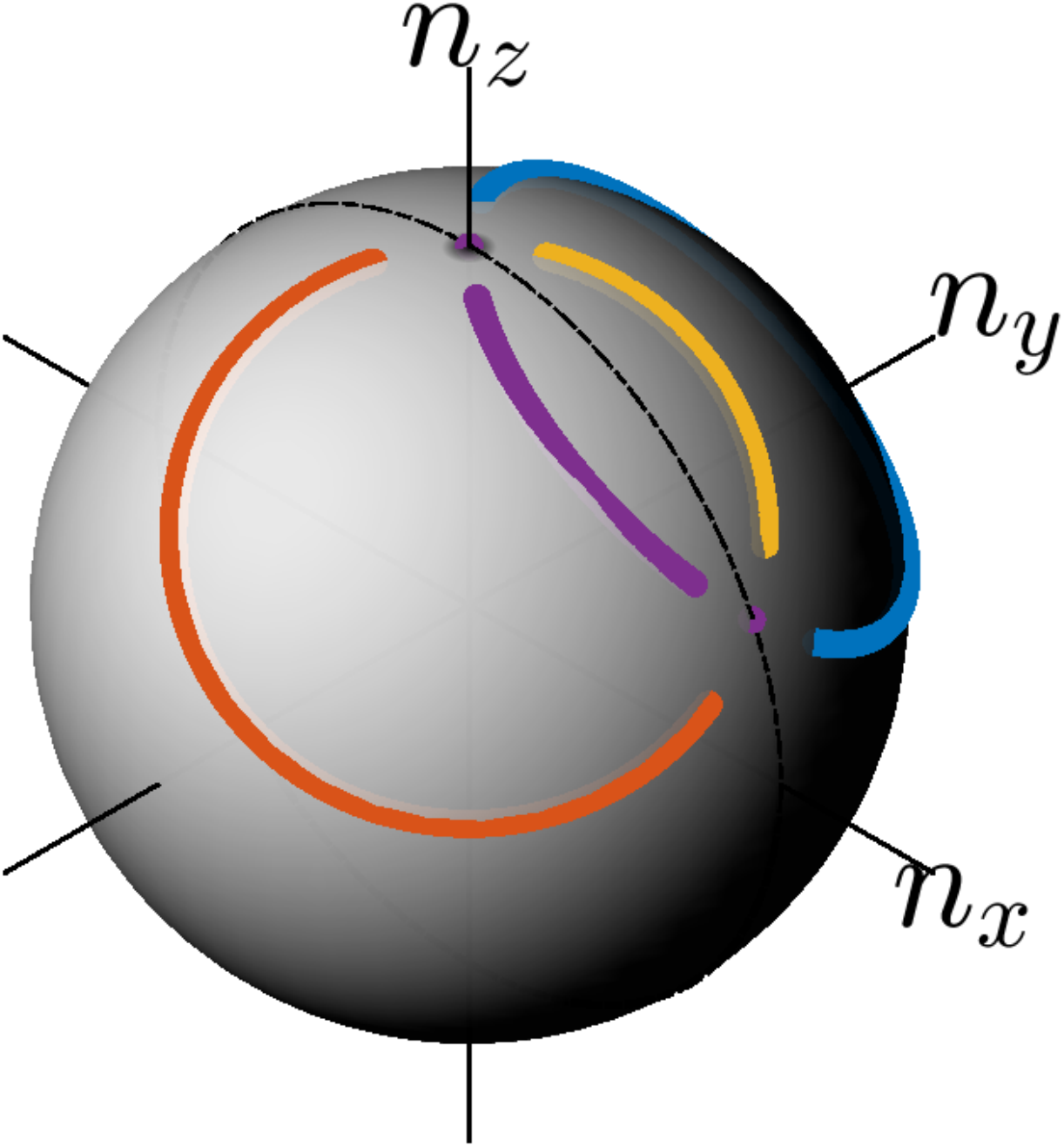}
			\label{fig:geo5D}}
		\caption{(Color online) Geometric decomposition of  geodesics in \subref{fig:geo4D} four- and \subref{fig:geo5D} five-dimensional state space for $\theta = \pi/3$. We can see that we obtain $n-1$ curves for $n$-dimensional system.}
		\label{fig:highergeo}
	\end{figure}

	We find that the $\vb{m}$ is same for any choice of the three Bloch vectors on the curve $\{\ket{\chi_k(s)}\}$. Moreover, the projection of all the Bloch vectors $\vb{p}(s)$ on the curve $\{\ket{\chi_k(s)}\}$ with $\vb{m}$ remains constant. This indicates that the curve traced by $ \ket{\chi_k(s)} $ is a circular segment for all values of $k$. The radius corresponding to the $k$th circular segment is 
	\begin{align}
		\label{eq:radius-geodesic}
		R_k = \dfrac{2 \beta}{\sqrt{4 \beta^2 - \alpha^2 (\Delta^* \omega_k^* - \Delta \omega_k)^2}}.
	\end{align}
	From the above expression, we see that the radii depend only on the inner product between the end states and not explicitly on the states. Furthermore, the circular segments are unique for the geodesic between a given set of end states. Therefore, once the end states are uniquely identified on the Bloch sphere, one can construct the desired geodesic by using Eq.~\eqref{eq:radius-geodesic} alone.  In Fig.~\ref{fig:highergeo}, we have plotted the structure of geodesic curve on the Bloch sphere for $n=4$- and $5$-dimensional state space. 
	
	There are certain intrinsic symmetries in the MS representation of a higher-dimensional geodesic. These symmetries reflect differently for even and odd dimensions. For example, if the dimension $n$ is odd, we get an even number of curves which appear in dual pairs, i.e., the pair of curves which are reflections of each other. The two curves $\{\ket{\chi_{i}(s)}\},\{\ket{\chi_{j}(s)}\}$, which are dual of each other, satisfy the condition $i + j = 0 \mod n-1 $, where $i,j = 0,1,2,\dots, n-2$, and $i\ne j$.  However, in the case of even $n$, one curve occurs along the great circle connecting the end states, and the remaining $n-2$ curves occur in dual pairs. For even $n$, the two dual curves $\{\ket{\chi_{i}(s)}\},\{\ket{\chi_{j}(s)}\}$ satisfy the condition $i + j = n-2 \mod n-1$. From Eqs.~\eqref{normal-vector} and \eqref{eq:radius-geodesic}, it is evident that the dual curves give the same value of radius with different centers.
	
	Now we have a way of constructing a unique geodesic between any two states $ \{\ket{\Psi_1}, \ket{\Psi_2}\} $ in the $n$-dimensional state space. Given $\ket{\Psi_1},~\ket{\Psi_2}$ we first map them to $(n-1)$-fold degenerate MSs states $\ket*{\tilde\Psi_1},~\ket*{\tilde\Psi_2}$ in the MS representation using a unitary transformation $U$. Thereafter, we construct circular segments on the Bloch sphere which are defined by Eqs.~\eqref{normal-vector} and~\eqref{eq:radius-geodesic} yielding the  geodesic in $n$-dimensional state space, connecting the states $ \ket*{\tilde{\Psi}_1} $ and $ \ket*{\tilde{\Psi}_2} $. Finally, we apply the $U^\dagger$  to get the desired geodesic between the original states.
	
	\section{Bloch Sphere decomposition of NPCs}\label{Sec:NPC}
	So far, we have seen that the geometric decomposition of the geodesics has revealed an interesting underlying symmetry. This has not only given us a better understanding of the geodesics but also provided us geometric ways of constructing one.  In this section, we investigate the geometric structure of NPCs in higher-dimensional state space. We mostly deal with the three-dimensional case, but the analysis can easily be extended to $n$-dimensional state space. Here, we propose a way to construct NPCs by choosing a suitable set of curves on the Bloch sphere. 
	
	Once again, we consider the  end states 
	$		\ket*{\tilde\Psi_1} = \ket{0} \otimes \ket{0}$ and 
	$\ket*{\tilde\Psi_2} = \ket{\phi} \otimes \ket{\phi}$ as defined earlier. To construct an NPC between $\ket{\Psi_1}$ and $\ket{\Psi_2}$ we propose the following:
	
	{\bf Proposition:} An arbitrary curve $\mathcal{C}$ connecting the states $\{\ket*{\tilde\Psi_1},\ket*{\tilde\Psi_2}\}$ and its dual curve $\mathcal{C}^*$  together form an NPC in the three-dimensional state space.
	
	{\it Proof}:
	The most general curve  $\mathcal{C} \equiv \{\ket{\psi(s)}|s_1\le s \le s_2\}$ connecting the states $\{\ket*{\tilde\Psi_1},\ket*{\tilde\Psi_2}\}$ is given by  
	\begin{equation}
		\label{NPC-curve1}
		\ket{\psi(s)} = \begin{pmatrix}
			\cos[\eta(s)/2]\\
			e^{i\Gamma(s)} \sin[\eta(s)/2]\\
		\end{pmatrix},
	\end{equation}
	where $\eta(s)$ and $\Gamma(s)$ are arbitrary real functions of $s$. The functions $\eta(s)$ and $\Gamma(s)$ satisfy the relations $\eta(s_1)=0,\eta(s_2)=2\cos^{-1}(\alpha)$ and $\Gamma(s_1)=\Gamma(s_2)=0$. The dual curve $\mathcal{C}^* \equiv \{\ket{\psi(s)}|s_1\le s \le s_2\}$ can then be defined as
	\begin{equation}
		\label{NPC-curve2}
		\ket{\psi'(s)}=\begin{pmatrix}
			\cos[\eta(s)/2]\\
			e^{-i\Gamma(s)} \sin[\eta(s)/2]\\
		\end{pmatrix}.
	\end{equation}
	Using the curves $\mathcal{C}$ and $\mathcal{C}^*$, the states on the curve in the three-dimensional state space can be written as
	\begin{equation}
		\label{Majorana-representation-NPC}
		\ket*{\tilde\Psi(s)}=\dfrac{1}{\mathcal{N}(s)}\big[\ket{\psi(s)}\ket{\psi'(s)}+\ket{\psi'(s)}\ket{\psi(s)}\big],
	\end{equation}
	where $\mathcal{N}(s)$ is the normalization constant. One can find that the BI of order three for the states in Eq.~\eqref{Majorana-representation-NPC} is always real and positive (see Appendix~\ref{Appendix:ThirdOrderBI}); hence, the curve $\{\ket*{\tilde\Psi(s)}\}$ is an NPC. For the given inner product between the end states, by choosing  $\eta(s)$ and $\Gamma(s)$, one can generate NPCs.
	Since one can construct infinitely many curves $\mathcal{C}$ connecting the states $\{\ket*{\tilde\Psi_1},\ket*{\tilde\Psi_2}\}$, we can construct infinitely many NPCs between these two states. 
	
	There is a particularly interesting subclass of  NPC which can be useful. An arc of a great circle passing through the initial and final states is dual to itself and referred as \emph{self-dual}.  This kind of curve on the Bloch sphere is given by (up to a unitary transformation)
	\begin{equation}
		\label{self}
		\ket{\psi(s)}=\begin{pmatrix}
			\cos[\eta(s)/2]\\
			\sin[\eta(s)/2]\\
		\end{pmatrix}; \qquad \,s_1\le s \le s_2,
	\end{equation}
	with $\eta(s_1)=0, \; \eta(s_2)=2\cos^{-1}(\alpha)$. A curve $\{\ket*{\tilde\Psi(s)}=\ket{\psi(s)}\otimes \ket{\psi(s)}\}$ is also an NPC, which we shall call a self-dual curve.
	
\begin{figure}
		\subfigure[]{
			\includegraphics[width=3.5cm]{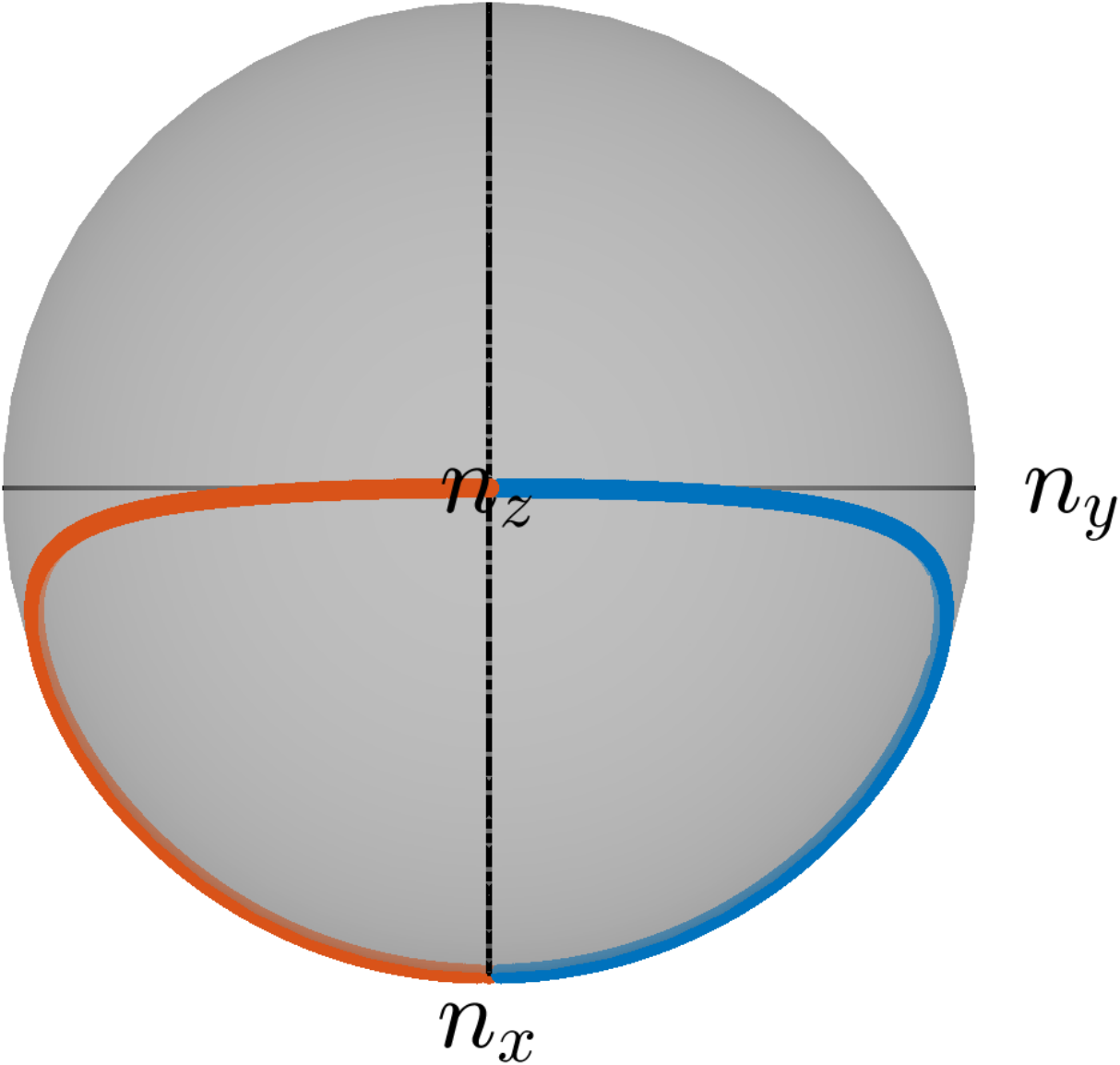}
			\label{fig:npc1}}
		\subfigure[]{
			\includegraphics[width=3.5cm]{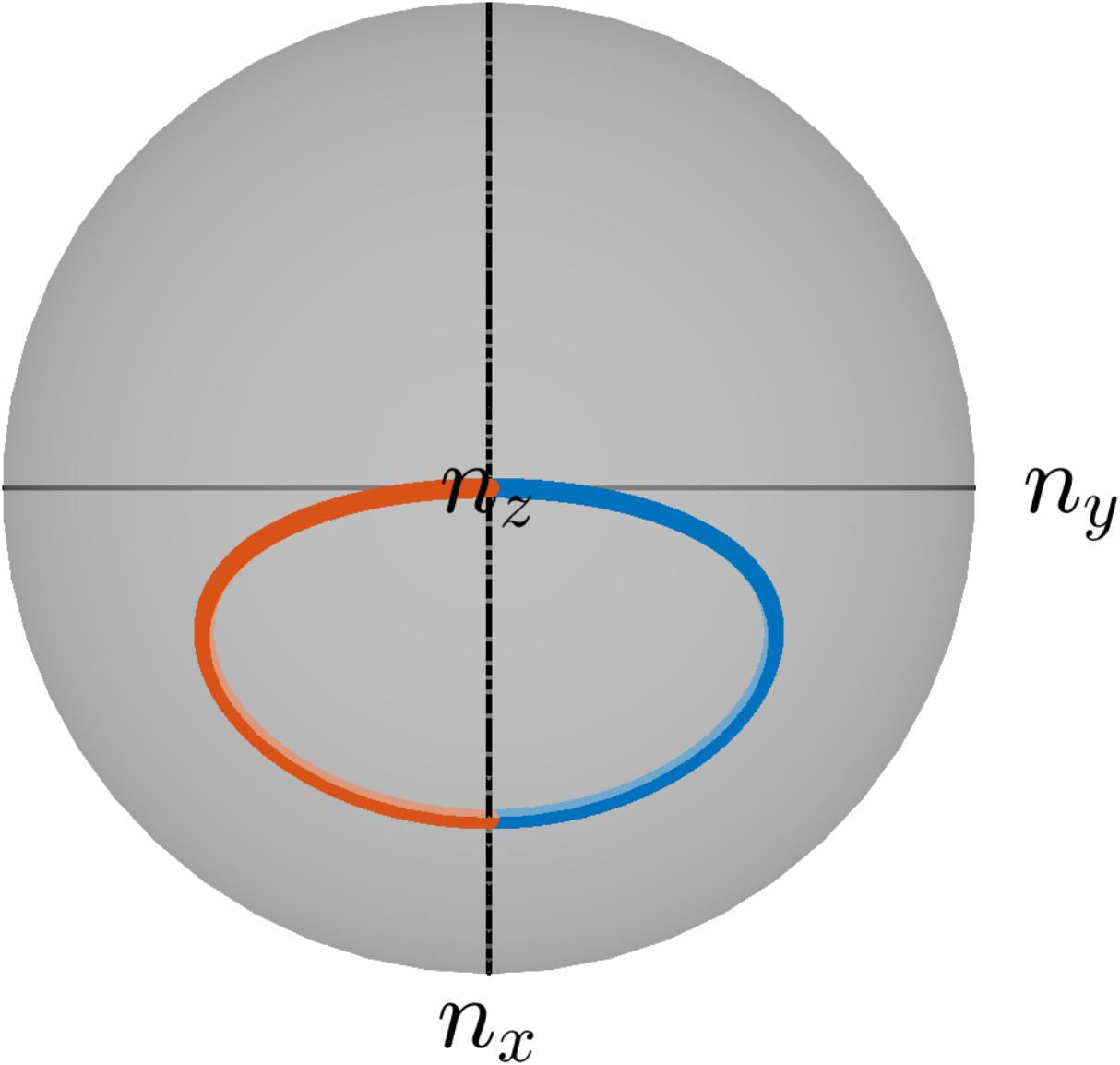}
			\label{fig:npc2}}
		\subfigure[]{
			\includegraphics[width=3.5cm]{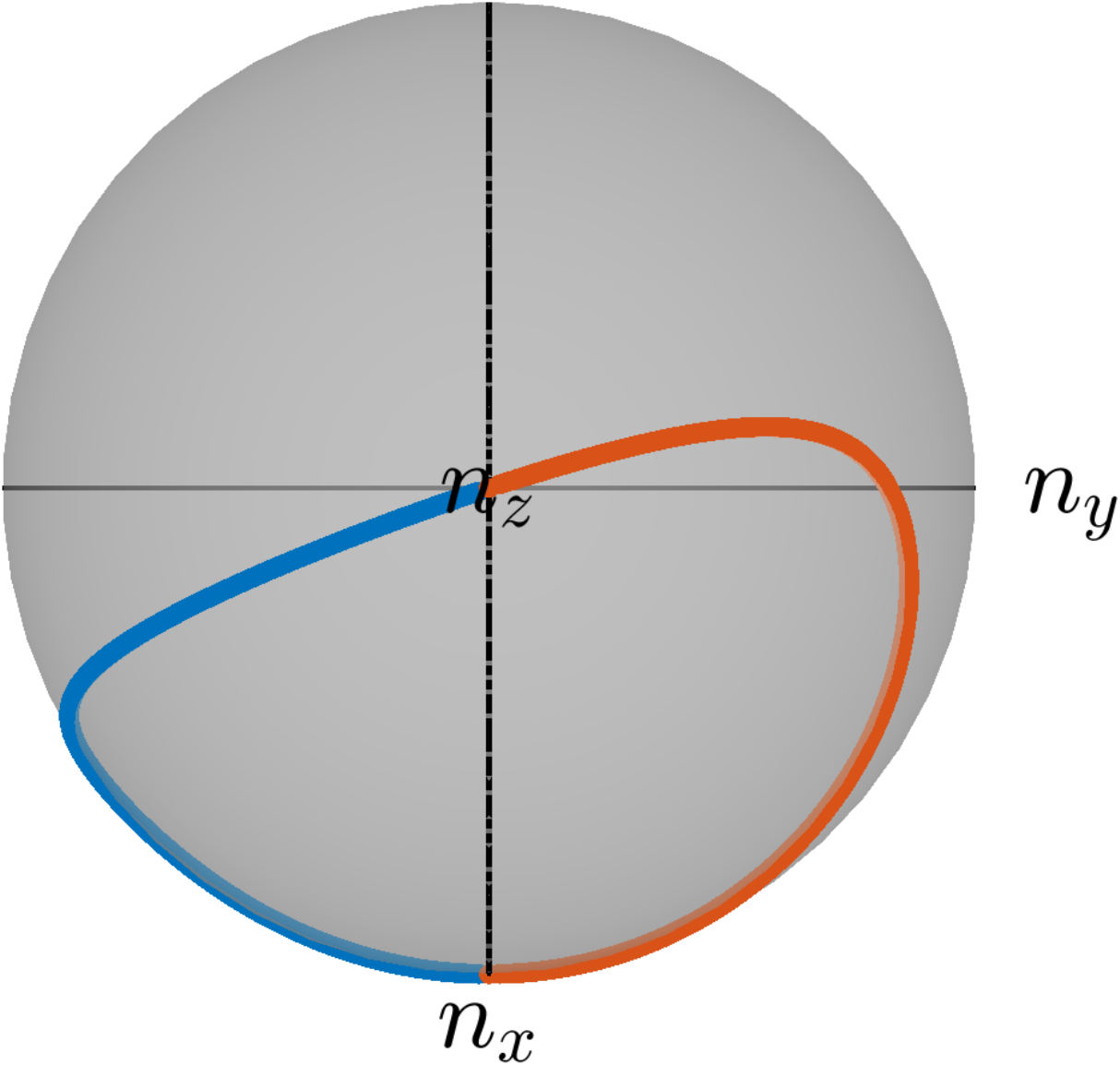}
			\label{fig:npc1b}}
		\subfigure[]{
			\includegraphics[width=3.5cm]{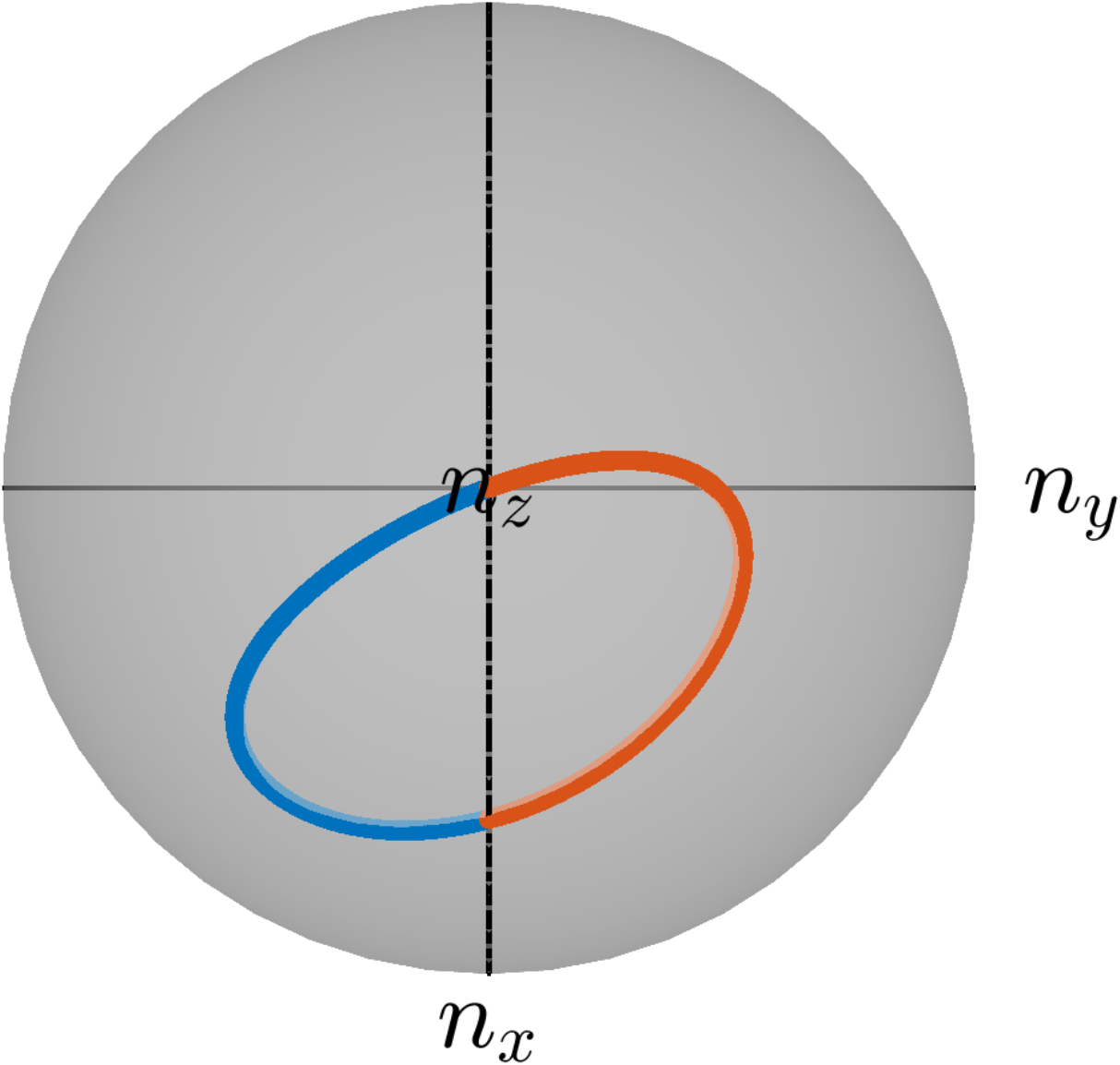}
			\label{fig:npc2b}}
		\caption{(Color online) Here, we plot the Bloch geometric decomposition of null-phase curves (NPCs) in three-dimensional state space between the two states given in Eq.~\eqref{Eq:39}. In \subref{fig:npc1} $\theta = \pi/3$, \subref{fig:npc2} $\theta = \pi/6$, we plot the NPCs which are constructed by considering dual pairs of curves, whereas in \subref{fig:npc1b} and \subref{fig:npc2b} we plot the NPCs which are constructed by geometric decomposition of the curve in Eq.~\eqref{eq:nullphasecurve-II}. We have chosen $\chi = \pi/3$ and the same values of $\theta$ as in \subref{fig:npc1} and \subref{fig:npc2}.}
		\label{fig:npc}
	\end{figure}

	Although, we have presented the construction of NPCs for the end states which can be represented by degenerate MSs, the same technique can be used to construct NPCs for arbitrary end states. As an example, let us construct an NPC connecting the two nonorthogonal states $\ket{\Psi_1} = (1 \; 0 \; 0)^T$ and $\ket{\Psi_2} = (\cos\theta \; \sin\theta \;\; 0)^T$ in three-dimensional state space. Since the state $\ket{\Psi_2}$ cannot be represented by degenerate MSs, we first apply a common unitary transformation $U$ to bring this to the state represented by degenerate MSs. The appropriate unitary transformation $U$ is of the form:
	\begin{align}
		U = \begin{pmatrix}
			1 & 0 & 0 \\
			0 & a & -b \\
			0 & b & a 
		\end{pmatrix},
	\end{align}
	where $a=\sqrt{2} \alpha \beta/\sin \theta $, $b= \beta^2/\sin \theta$, $\alpha^2 = \cos \theta$, and $\alpha^2 + \beta^2 = 1$. The states after applying $U$ read
	
	\begin{align}\label{Eq:39}
		\ket*{\Psi'_1} = \begin{pmatrix}
			1 \\
			0 \\
			0
		\end{pmatrix}, \qquad	\ket*{\Psi'_2} =  \begin{pmatrix}
			\cos \theta \\
			a \sin \theta \\
			b\sin \theta
		\end{pmatrix},
	\end{align}
	which can further  be written as 
	\begin{align*}
		\ket*{\tilde{\Psi}_1} &= \ket{0} \otimes \ket{0}, \\
		\ket*{\tilde{\Psi}_2} &= \ket{\phi} \otimes \ket{\phi}, \\
	\end{align*}
	where $\ket{0}$ and $ \ket{\phi} $ are the same as defined earlier. We now take a pair of dual curves given in Eqs.~\eqref{NPC-curve1} and \eqref{NPC-curve2} with appropriate boundary conditions on $\eta(s)$ and $\Gamma(s)$. Consequently, the state in three-dimensional state is written as 
	\begin{align*}
		\ket*{\Psi'} = \mathcal{N} \begin{pmatrix}
			2 \cos^2[\eta(s)/2] \\
			\sqrt{2} \cos[\eta(s)/2] \sin[\eta(s)/2] \cos[\Gamma(s)] \\
			2 \sin^2[\eta(s)/2]
		\end{pmatrix},
	\end{align*}  
	where $\mathcal{N}$ is the normalization constant. Now we apply $U^{\dagger}$ on $ \ket*{\Psi'} $ to get the state $\ket{\Psi}$ back. With the appropriate choice of $\eta$ and $\Gamma$,  we can bring it to the form which is already given in Ref.~\cite{Arvind2003} as an example of an NPC. Further, we choose the functions $\eta(s)$ and $\Gamma(s)$ with $0\le s \le \theta$ as
	\begin{align}
		\eta(s)&=\cos^{-1}\Big[\frac{A(s)-C(s)}{A(s)+C(s)}\Big], \nonumber \\
		\Gamma(s)&=\tan^{-1}\Big[\frac{\sqrt{4 A(s) C(s) - B^2(s)}}{B(s)}\Big],
	\end{align}
	where
	\begin{align}
		A(s) &= g(s) \cos s, \nonumber\\
		B(s) &= b\left[1 - g(s)^2\right]^{1/2}-a g(s) \sin s,\nonumber\\
		C(s) &= a  \left[1 - g(s)^2\right]^{1/2} + b g(s) \sin s,
	\end{align}
	with $0\le g(s)\le 1$, and $g(0)=g(\theta)=1$. This particular choice of the functions results in an NPC given by
	\begin{align} \label{eq:nullphasecurve-I}
		\ket{\Psi(s)} = \begin{pmatrix}
			g(s) \cos s  \\
			g(s) \sin s  \\
			\left[1 - g(s)^2\right]^{1/2}
		\end{pmatrix};\qquad 0\le s \le \theta,
	\end{align}
	which is the same derived in Ref~\cite{Arvind2003} between the states $\ket{\Psi_1} = (1 \; 0 \; 0)^T$ and $\ket{\Psi_2} = (\cos\theta \; \sin\theta \;\; 0)^T$. In Figs.~\ref{fig:npc1},~\ref{fig:npc2} we plot the geometric construction of NPCs on the Bloch sphere where we have chosen  $g(s)=\cos [s(s-\theta)]$.
	
	We have provided  methods to generate NPCs in three-dimensional state space using pairs of dual curves on the Bloch sphere. Our method of constructing NPCs can be easily extended to $n$-dimensional state space. For example, for odd $n$, we can construct $(n-1)/2$ pairs of dual curves to get an NPC. For the case when $n$ is even, we take $(n-2)/2$ pairs of dual curves and one curve  along the great circle connecting the end states, to construct the NPC. 
	
	{\it An example of NPCs which do not come under this category:}     
	Interestingly, there exist NPCs connecting the considered end states $\ket{\Psi_1}$ and $\ket{\Psi_2}$ which cannot be constructed either by a pair of dual curves or by self-dual curves. These NPCs are obtained from Eq.~\eqref{eq:nullphasecurve-I} by applying a unitary transformation which reads
	\begin{align}
		V = \begin{pmatrix}
			1 & 0 & 0 \\
			0 & 1 & 0  \\
			0 & 0 & e^{i \chi} 
		\end{pmatrix}.
	\end{align}
	The BI is invariant under such unitary transformations and results in another NPC. The new NPC reads
	\begin{equation} \label{eq:nullphasecurve-II}
		\ket{\Psi(s)} \mapsto V \ket{\Psi(s)} = \begin{pmatrix}
			g(s) \cos s  \\
			g(s)\sin s  \\
			e^{i\chi} \left[1 - g(s)^2\right]^{1/2} 
		\end{pmatrix},
	\end{equation}
	where the parameter $0\le s \le \theta$, $\chi \in \mathbb{R}$ is a constant phase factor, $0\le g(s)\le 1$, and $g(0)=g(\theta)=1$. After a common unitary transformation, these NPCs decompose to a pair of curves which are not reflective about the great circle connecting the degenerate MSs on the Bloch sphere as shown in Figs.~\ref{fig:npc1b} and \ref{fig:npc2b}. Therefore, this kind of NPC cannot be constructed by choosing $\eta(s)$ and $\Gamma(s)$.

	\section{Conclusion}\label{sec:Conclusion}
	
	In conclusion, we have developed a Bloch sphere geometric decomposition for geodesics and NPCs in higher-dimensional state space.  We have shown that geodesics connecting the two $(n-1)$-fold degenerate MS states in $n$-dimensional state space decompose to $n-1$ circular segments on the Bloch sphere. The circular segments, which are uniquely determined by the radius and the end states, can be completely characterized by the MSs corresponding to the end states and the inner product between them. Therefore, once the MSs for the end states are known, we can construct the geodesic between any given states in $n$-dimensional state space. 
	We have also shown that NPCs can be constructed between any given end states by constructing arbitrary pairs of dual curves between the end states. 
	A particularly interesting class of NPC is the one where all the curves are dual of themselves, i.e., self-dual curves. The geometric decomposition presented in this paper reveals intrinsic symmetries in the geodesics and NPCs and improves our understanding of the geometric structure of the higher-dimensional state space.

	\begin{acknowledgments}
		S.K.G. and V.M. acknowledge the financial support from Interdisciplinary Cyber Physical Systems (ICPS) programme of the Department of Science and Technology, India (Grant No. DST/ICPS/QuST/Theme-1/2019/12)
	\end{acknowledgments}
	
	\appendix
	\section{Transforming two arbitrary states to degenerate MSs simultaneously}
	\label{AppendixA}
	Consider two arbitrary states $ \{\ket{\Psi_1}, \ket{\Psi_2}\} $ in three-dimensional state space with the inner product given by $\ip*{\Psi_1}{\Psi_2} = \cos \theta$, i.e., real and positive. We can write $ \ket{\Psi_2} $ as
	\begin{align}
		\ket{\Psi_2} = \cos\theta \ket{\Psi_1} + \gamma \ket*{\bar{\Psi}_1}
	\end{align}
	where $\gamma = \sin\theta e^{i \phi}$, $\phi \in \mathbb{R}$ and $\ket*{\bar{\Psi}_1}$ is orthogonal to $\ket{\Psi_1}$. Now, we take unitary of the form 
	\begin{align}
		U =\dyad{0}{\Psi_1} + e^{-i \phi} \dyad*{1}{\bar{\Psi}_1} + \dyad*{2}{\bar{\bar{\Psi}}_1}
	\end{align} 
	where $\{\ket{0}, \ket{1}, \ket{2}\}$ and $\{ \ket{\Psi_1}, \ket*{\bar{\Psi}_1}, \ket*{\bar{\bar{\Psi}}_1}\}$ form the orthonormal bases. On application of this unitary on the two states $ \{\ket{\Psi_1}, \ket{\Psi_2}\} $ results in the states of the form
	\begin{align} \label{eq:general-states}
		U\ket{\Psi_1} &\equiv \ket{\Psi'_1}= \begin{pmatrix}
			1 \\
			0 \\
			0
		\end{pmatrix},\\ 
		U\ket{\Psi_2} &\equiv \ket{\Psi'_2}= \begin{pmatrix}
			\cos\theta \\
			\sin\theta \\
			0
		\end{pmatrix}.
	\end{align} 
	One can see that $\ket{\Psi'_2}$ cannot be represented by a degenerate MSs. For that, one needs to apply one more unitary transformation to bring $\ket{\Psi'_2}$ to a state represented by degenerate MSs. An appropriate unitary for that is given by
	\begin{align} \label{eq:unitaryU}
		\tilde{U} = \begin{pmatrix}
			1 & 0 & 0 \\
			0 & a & -b \\
			0 & b & a 
		\end{pmatrix},
	\end{align}
	where $a=\sqrt{2} \alpha \beta/\sin \theta $, $b= \beta^2/\sin \theta$, $\alpha^2 = \cos \theta$, and $\alpha^2 + \beta^2 = 1$. After applying the following unitary, we will get
	\begin{align}
		\ket*{\Psi_2} = \begin{pmatrix}
			\alpha^2 \\
			\sqrt{2} \alpha \beta \\
			\beta^2
		\end{pmatrix},
	\end{align}
	which is written as 
	\begin{align}
		\ket*{\tilde\Psi_2} = \begin{pmatrix}
			\alpha \\
			\beta
		\end{pmatrix} \otimes\begin{pmatrix}
			\alpha \\
			\beta
		\end{pmatrix}.
	\end{align}
	Therefore, any two arbitrary states with a real inner product can be brought to the degenerate MSs states. 
	
	Similarly given any two arbitrary states $ \{\ket{\Psi_1}, \ket{\Psi_2}\} $ in $ n $-dimensional state space with the inner product $\ip*{\Psi_1}{\Psi_2} = \cos \theta$, we can always write $\ket{\Psi_2}$ as~\cite{Akhilesh2020}
	\begin{align}
		\ket{\Psi_2} = \cos\theta \ket{\Psi_1} + \gamma \ket*{\bar{\Psi}_1}
	\end{align}
	where $\gamma = \sin\theta e^{i \phi}$, $\phi \in \mathbb{R}$ and $\ket*{\bar{\Psi}_1}$ is orthogonal to $\ket{\Psi_1}$.
	\\
	Let us take a unitary $V$ of the form
	\begin{align} \label{eq:unitrayV}
		V =\dyad{0}{\Psi_1} + e^{-i \phi} \dyad*{1}{\bar{\Psi}_1} +\dyad*{2}{\bar{\Psi}_2} \dots +  \dyad*{n-1}{\bar{\Psi}_{n-1}}
	\end{align} 
	where $\{\ket{0}, \ket{1}, \dots, \ket{n-1}\}$ and $\{ \ket{\Psi_1}, \ket*{\bar{\Psi}_1}, \dots, \ket*{\bar{\Psi}_{n-1}}\}$ are the two sets of orthonormal bases. Using this unitary, the two states can be brought to the following form:
	\begin{align} \label{eq:general-statesd}
		\ket{\Psi'_1}= \begin{pmatrix}
			1 \\
			0 \\
			0 \\
			\vdots \\
			0
		\end{pmatrix}, \qquad \ket{\Psi'_2}= \begin{pmatrix}
			\cos\theta \\
			\sin\theta \\
			0\\
			\vdots \\
			0
		\end{pmatrix}.
	\end{align}
	Further, the state $\ket*{\Psi'_2}$ can be transformed to a state represented by degenerate MSs states by applying a suitable unitary matrix [like the one given in Eq.~\eqref{eq:unitaryU} for the three-dimensional case] without changing the form of the state $\ket*{\Psi'_1}$~\cite{Tamate2011}. The general form of the unitary matrix is given as 
	\begin{align} \label{eq:rotational_matrix}
		\tilde{V} = \left(\begin{array}{c|c c c}
			1 & 0 & \cdots & 0\\
			\hline
			0 & \\
			\vdots &   &  R & \\
			0
		\end{array} \right)
	\end{align}
	The first column of the block matrix $R$ is taken to be
	\begin{align} \label{eq:column_R}
		\dfrac{1}{\sin \theta }\begin{pmatrix}
			\sqrt{^{n-1}C_1}\alpha^{n-2}\beta\\
			\sqrt{^{n-1}C_2}\alpha^{n-3}\beta^2\\
			\vdots\\
			\beta^{n-1}
		\end{pmatrix}.
	\end{align}
	The other columns of the block matrix $R$ can be constructed which satisfying the following condition
	\begin{align} \label{eq:rotation_column_equation}
		y_1\left(\sqrt{^{n-1}C_1}\alpha^{n-2}\beta\right)+ \dots + y_{n-1} \left(\beta^{n-1}\right) = 0,
	\end{align}
	where $y^T=(y_1, y_2, y_3, \cdots, y_{n-1})$ represents the constructed columns of the block matrix $R$.
	For example, in the case of four-dimensional state space, the constructed unitary matrix will be of the form 
	\begin{align}
		\begin{pmatrix}
			1 & 0 & 0 & 0\\
			0 & \frac{\sqrt{3}\alpha^2 \beta}{\sin\theta} & -\beta & \frac{-\alpha\beta^2}{\sqrt{3\alpha^2+\beta^4}}\\
			0 & \frac{\sqrt{3}\alpha \beta^2}{\sin\theta} & \alpha & \frac{-\beta^3}{\sqrt{3\alpha^2+\beta^4}}\\
			0 & \frac{ \beta^3}{\sin\theta} & 0 & \frac{\sqrt{3}\alpha}{\sqrt{3\alpha^2+\beta^4}}
		\end{pmatrix}.
	\end{align}
	This unitary matrix transforms $\ket*{\Psi'_2}$ to the state represented by degenerate MSs in the Majorana representation. 
	\section{Third order BI} \label{Appendix:ThirdOrderBI}
	The BI of third order $\Delta_3$ for any given three mutually non-orthogonal states $ \{\ket{\Psi_1}, \ket{\Psi_2}, \ket{\Psi_3}\} $ is written as
	\begin{align}
		\Delta_3(\Psi_1, \Psi_2, \Psi_3) = \ip*{\Psi_1}{\Psi_2}\ip*{\Psi_2}{\Psi_3}\ip*{\Psi_3}{\Psi_1}.
	\end{align}
	In the MR representation, these states are written as 
	\begin{align}
		\ket{\Psi_1}&=\frac{1}{\mathcal{N}_1}\left[\ket{\psi_1}\ket{\psi'_1}+\ket{\psi'_1}\ket{\psi_1} \right]\nonumber\\
		\ket{\Psi_2}&=\frac{1}{\mathcal{N}_2}\left[\ket{\psi_2}\ket{\psi'_2}+\ket{\psi'_2}\ket{\psi_2} \right]\nonumber\\
		\ket{\Psi_3}&=\frac{1}{\mathcal{N}_3}\left[\ket{\psi_3}\ket{\psi'_3}+\ket{\psi'_3}\ket{\psi_3} \right]
	\end{align}
	where $\mathcal{N}_i$ are the normalization constants. Using this representation, we can expand $ \Delta_3(\Psi_1, \Psi_2, \Psi_3) $ as follows
	\begin{align*}
		&\ip*{\Psi_1}{\Psi_2}\ip*{\Psi_2}{\Psi_3}\ip*{\Psi_3}{\Psi_1} \\
		&\quad= \ip*{\psi_1}{\psi_2}\ip*{\psi_2}{\psi_3}\ip*{\psi_3}{\psi_1}  \ip{\psi'_1}{\psi'_2} \ip{\psi'_2}{\psi'_3} \ip{\psi'_3}{\psi'_1} \\
		&\qquad + \ip*{\psi_1}{\psi_2}\ip*{\psi_2}{\psi'_3}\ip{\psi'_3}{\psi'_1} \ip{\psi'_1}{\psi'_2} \ip{\psi'_2}{\psi_3} \ip*{\psi_3}{\psi_1}   \\
		&\qquad+ \ip*{\psi_1}{\psi'_2}\ip{\psi'_2}{\psi'_3}\ip{\psi'_3}{\psi'_1} \ip{\psi'_1}{\psi_2}\ip*{\psi_2}{\psi_3} \ip*{\psi_3}{\psi_1}  \\ 
		&\qquad+ \ip*{\psi_1}{\psi'_2}\ip{\psi'_2}{\psi_3}\ip*{\psi_3}{\psi_1} \ip{\psi'_1}{\psi_2}\ip*{\psi_2}{\psi'_3}  \ip{\psi'_3}{\psi'_1} \\
		&\qquad+ \ip*{\psi_1}{\psi_2}\ip*{\psi_2}{\psi_3}\ip*{\psi_3}{\psi'_1} \ip{\psi'_1}{\psi'_2} \ip{\psi'_2}{\psi'_3}  \ip{\psi'_3}{\psi_1} \\
		&\qquad + \ip*{\psi_1}{\psi_2}\ip*{\psi_2}{\psi'_3}\ip{\psi'_3}{\psi_1} \ip{\psi'_1}{\psi'_2} \ip{\psi'_2}{\psi_3} \ip*{\psi_3}{\psi'_1}   \\
		&\qquad+ \ip*{\psi_1}{\psi'_2}\ip{\psi'_2}{\psi'_3}\ip{\psi'_3}{\psi_1} \ip{\psi'_1}{\psi_2}\ip*{\psi_2}{\psi_3} \ip*{\psi_3}{\psi'_1}  \\ 
		&\qquad+ \ip*{\psi_1}{\psi'_2}\ip{\psi'_2}{\psi_3}\ip*{\psi_3}{\psi'_1} \ip{\psi'_1}{\psi_2}\ip*{\psi_2}{\psi'_3}  \ip{\psi'_3}{\psi_1} 
	\end{align*}
	Now we choose
	\begin{align}
		\ket{\psi(s)} = \begin{pmatrix}
			\cos[\eta(s)/2]\\
			e^{i\Gamma(s)} \sin[\eta(s)/2]\\
		\end{pmatrix},  
	\end{align}
	and 
	\begin{align}
		\ket{\psi'(s)} = \begin{pmatrix}
			\cos(\eta(s)/2)\\
			e^{-i\Gamma(s)} \sin(\eta(s)/2)\\
		\end{pmatrix}.
	\end{align}
	It is very clear from the above choice that a pair of inner products $\ip*{\psi_i}{\psi_j}$ and $\ip*{\psi'_i}{\psi'_j}$, $\ip*{\psi_i}{\psi'_j}$ and $\ip*{\psi'_i}{\psi_j}$, or $\ip*{\psi_i}{\psi'_j}$ and $\ip*{\psi'_i}{\psi_j}$ are complex conjugate of each other. The $\Delta_3(\Psi_1, \Psi_2, \Psi_3)$ has eight terms where each term contains three such pairs. Hence, it is very straightforward to show that $\Delta_3(\Psi_1, \Psi_2, \Psi_3)$ is real and positive for the above choice of $ \ket{\psi(s)} $ and $ \ket{\psi'(s)} $.

\end{document}